\documentclass[aps,prb,preprint,showpacs,groupedaddress]{revtex4}

\input{tcilatex}

\begin{document}

\title{Effect of Hydrostatic Pressure on the Superconductivity in Na$_{x}$%
CoO $_{2}\cdot $yH$_{2}$O}
\author{B. Lorenz$^{1}$, J. Cmaidalka$^{1}$, R. L. Meng$^{1}$, and C. W. Chu$
^{1,2,3}$}
\affiliation{$^{1}$Department of Physics and TCSAM, University of Houston, Houston, TX
77204-5932}
\affiliation{$^{2}$Lawrence Berkeley National Laboratory, 1 Cyclotron Road, Berkeley, CA
94720}
\affiliation{$^{3}$Hong Kong University of Science and Technology, Hong Kong, China}
\date{\today }

\begin{abstract}
The effect of hydrostatic pressure on the superconducting transition
temperature of Na$_{0.35}$CoO$_{2}\cdot $yH$_{2}$O was investigated by ac
susceptibility measurements up to 1.6 GPa. The pressure coefficient of T$%
_{c} $ is negative and the dependence T$_{c}$(p) is nonlinear over the
pressure range investigated. The magnitude of the average dlnT$_{c}$/dp $%
\approx$ -0.07 GPa$^{-1}$ is comparable to the pressure coefficient of
electron-doped high-T$_{c}$ copper oxide superconductors with a similar
value of T$_{c}$. Our results provide support to the assumption of
two-dimensional superconductivity in Na$_{0.35}$CoO$_{2}\cdot $yH$_{2}$O,
which is similar to the cuprate systems, and suggest that intercalation of
larger molecules may lead to an enhancement of T$_{c}$.
\end{abstract}

\maketitle











The recent discovery of superconductivity in two-dimensional CoO$_{2}$
layers in the compound Na$_{0.35}$CoO$_{2}\cdot $yH$_{2}$O has attracted
attention because of its possible similarity to the high-T$_{c}$
superconductivity in CuO$_{2}$-planes of the cuprate systems but with a
different spin state associated with Co.\cite{1} The compound appears to be
the first superconducting layered metal oxide involving 3d-transition metals
other than copper. The de-intercalation of Na$^{+}$ ions accompanied by an
intercalation of water molecules results in a combined doping and c-axis
expansion effect effectively reducing the interlayer coupling between the CoO%
$_{2}$ planes and enhancing the two-dimensional character of the structure.
Magnetic susceptibility measurements have shown a diamagnetic drop below 5 K
ascribed to the onset of superconductivity although the estimated volume
fraction is only of the order of 10 \% of the value for perfect
diamagnetism. The high-field dependence of the susceptibility resembles that
of the high-T$_{c}$ copper oxides, with little change of the onset but an
appreciable broadening of the superconducting transition under increasing
magnetic field. Because of these similarities and the fact that the removal
of water suppresses the superconductivity it was suggested that the large
separation of the CoO$_{2}$ layers and the reduced interaction between them
is essential for stabilizing the superconducting state.\cite{1}

The major difference to the square CuO$_{2}$ planes of the cuprates is the
triangular symmetry of the Co-lattice that may lead to possible frustration
effects in the magnetic structure of the antiferromagnetically interacting
cobalt spins. Theoretical treatments therefore employ the t-J model on a
planar triangular lattice.\cite{2,3,4} Different magnetic and
superconducting orders, e.g. d-wave superconductivity,\cite{2} a
time-reversal-symmetry-breaking $d_{x^{2}-y^{2}}+id_{xy}$ ground state,\cite%
{3} and spin triplet superconductivity\cite{4} have been proposed. To
distinguish between different theoretical proposals and to facilitate our
basic understanding of superconductivity in Na$_{0.35}$CoO$_{2}\cdot $yH$_{2}
$O far more experimental work needs to be conducted. In particular, the
dependence of the superconducting state on the interlayer coupling is of
interest. The CoO$_{2}$ interlayer coupling should be tuned reversibly by
changing the distance between the planes (c-axis lattice parameter) without
affecting the chemical composition, i.e. the content of sodium and water.
This can be achieved by applying hydrostatic pressure. Because of the strong
lattice anisotropy and the weak bonding between the layers (the water
molecules are electrically neutral although polarized) it is expected that
the compression of the c-axis increases the interlayer coupling and has a
major effect on the superconductivity.

We have measured the superconducting T$_{c}$ of Na$_{0.35}$CoO$_{2}\cdot $yH$%
_{2}$O powder by ac magnetic susceptibility experiments in a hydrostatic
pressure environment up to 1.6 GPa. The observed decrease in T$_{c}$ of up
to 10 \% at 1.6 GPa indicates the importance of a reduced\ interlayer
coupling for stabilizing the superconducting ground state in this compound.

The Na$_{x}$CoO$_{2}\cdot $yH$_{2}$O powder was synthesized as described in
Ref. 1 and 5. The x-ray spectrum shows the reflections of the hexagonal
space group P6$_{3}/$mmc with lattice parameters a=2.820 \AA\ and c=19.593
\AA . The structural parameters are in good agreement with the recently
reported data.\cite{1} Magnetic susceptibility measurements indicate the
onset of superconductivity at 4.7 K, the magnitude of the diamagnetic signal
at 2 K in field and zero field cooling runs are comparable with data of Ref.
1. Before starting the high pressure experiment the sample was carefully
checked with respect to chemical compatibility with the liquid pressure
transmitting medium (a 1:1 mixture of 3M Fluorinert FC70/FC77). The pressure
medium was mixed with the sample powder and the dc susceptibility of the
mixture was measured using the SQUID magnetometer (Quantum Design). No
change or degradation in the superconducting properties of the mixture as
compared to the original powder was detected. For high-pressure ac
susceptibility measurements a dual coil was mounted onto a plastic frame
with a 1.3 mm diameter, 5 mm long concentric inner space accommodating the
sample powder and a small piece of high purity (99.9999 \%) lead for in situ
pressure measurements. The superconducting transition of the lead induced a
small diamagnetic signal in the pickup coil and the pressure was determined
from the shift of the lead transition temperature.\cite{9} The coil was
inserted into a Teflon container filled with the Fluorinert pressure medium
and pressure was generated by a beryllium-copper piston cylinder clamp.\cite%
{6} The pressure cell was inserted into a $^{4}$He dewar that allowed the
control of temperature between room temperature and 1.2 K. The temperature
below 45 K was measured by a germanium resistor built into the pressure
clamp close to the sample position. The inductance was measured with an LR
700 mutual inductance bridge (Linear Research) at a frequency of 19 Hz.

The inductance, I(T), measured at different pressures is shown in Fig. 1
(for clarity not all sets of acquired data are shown). The superconducting
transition is well resolved and the onset of diamagnetism causes the
decrease of the inductance signal below T$_{c}$. The vertical drop of \ I(T)
close to 7 K in Fig. 1 indicates the diamagnetic signal from the
superconducting transition of the lead manometer and it was used for
pressure determination. With increasing pressure the superconducting
transition of the Na$_{0.35}$CoO$_{2}\cdot $yH$_{2}$O powder shifts to lower
temperature and the transition becomes slightly sharper. It is remarkable
that the shape and the overall drop of the susceptibility curves do not
change appreciably with pressure which indicates that the sample is stable
at the applied hydrostatic pressure. There is no sign of a possible chemical
decomposition or loss of water during the experiments. After acquiring the
highest pressure data the pressure was almost completely released (a
residual pressure of 0.04 GPa was measured) and the susceptibility curve was
found in good agreement with the ambient pressure data taken at the start of
the experiments. These data are marked by filled squares in Fig. 1. This
again shows that the sample is stable and the superconducting properties are
reversible under hydrostatic pressure. The T$_{c}$(p) as estimated from the
onset of the diamagnetic signal (see inset of Fig. 1) is displayed in Fig.
2. For low pressure the change of T$_{c}$ is relatively small. With
increasing pressure the T$_{c}$ decreases faster resulting in a non linear
dependence T$_{c}$(p) as shown in Fig. 2. The solid line is a fit of the
data to a parabolic pressure dependence, $T_{c}(p)=4.68-0.0047\ p-0.183\
p^{2}$ (T$_{c}$ in K and p in GPa). At the highest pressure of this
experiment (1.53 GPa) the superconducting transition dropped to 4.2 K.
Extrapolating the parabolic dependence to T$_{c}$=0 yields a critical
pressure of $p_{c}\simeq 5$ GPa. The recovery of the T$_{c}$ after pressure
release is indicated by the filled square in Fig. 2 (in contrast to the open
circles denoting data estimated within the increasing pressure cycle). It
should be noted that in the present experiments the pressure medium was
frozen at the superconducting transition temperature. This may result in
deviations from pure hydrostatic conditions and small pressure gradients
throughout the sample space could be expected. Any pressure medium, even He
gas at elevated pressure, will solidify at these low temperatures. In order
to check on possible pressure gradients we have evaluated the width of the
diamagnetic drop of the ac susceptibility in passing the superconducting
transition of the lead manometer. The cooling speed was electronically
controlled as low as 0.01 K/min. The width of the lead transition (as
measured between the 90 \% and 10 \% drop of the susceptibility) was even at
the highest pressure still below 0.01 K and only slightly larger than the
zero pressure value of 0.008 K. If the small difference is attributed to a
pressure gradient across the volume of the lead sample the inhomogeneity is
less than 0.005 GPa at the maximum pressure of 1.6 GPa. With the Na$_{0.35}$%
CoO$_{2}\cdot $yH$_{2}$O samples linear dimension being about 10 times the
size of the lead the possible pressure gradients are still small and should
not affect the results significantly.

A negative pressure coefficient of T$_{c}$ is frequently observed in low- as
well as high-T$_{c}$ compounds. In low-T$_{c}$ (i.e. BCS-like,
phonon-mediated) superconductivity the negative pressure shift of T$_{c}$
can be explained by a decrease of the electron-phonon coupling constant, $%
\lambda $. $\lambda $ is proportional to the electronic density of states
(DOS) and to the inverse of the average square of the phonon frequency. The
application of pressure usually results in an increase of the phonon
frequency (phonon ''hardening'' effect) and, in many instances, in a
decrease of the DOS. Both effects lead to a decrease of T$_{c}$, as was
recently discussed, for example, in MgB$_{2}$.\cite{10} The high-T$_{c}$
cuprate superconductors exhibit both, positive as well as negative pressure
coefficients, depending on the compound and the doping state. For some
hole-doped copper oxides a negative dT$_{c}$/dp was reported in the
overdoped regime and partially attributed to a pressure-induced charge
transfer but it was also shown that this charge transfer is not the only
factor affecting the pressure shift of T$_{c}$ (for an extensive discussion
see Ref. 9). Na$_{0.35}$CoO$_{2}\cdot $yH$_{2}$O is, however, an
electron-doped system. This has to be taken into account if the present data
are compared with the high-T$_{c}$ cuprates. For electron-doped cuprates,
e.g Sm$_{1.88}$Ce$_{0.12}$CuO$_{4-y}$, the pressure coefficient of T$_{c}$
was found to be negative for a large number of systems with different
chemical composition and (electron) doping.\cite{7,8} Thereby, a universal
relation between the relative pressure coefficient, dlnT$_{c}$/dp, and T$%
_{c} $ was proposed.\cite{8} According to this relation the pressure
coefficient for a sample with T$_{c}$=4.7 K should be of the order of -0.1
to -0.15 GPa$^{-1}$. The T$_{c}$-p relation shown in Fig. 2 is nonlinear,
however, estimating the average pressure change of T$_{c}$ yields a value of
dlnT$_{c} $/dp=-0.07 GPa$^{-1}$and calculating the slope of T$_{c}$(p) at
the high pressure end of the graph we get dlnT$_{c}$/dp=-0.1 GPa$^{-1}$.
These values are in fair agreement with the typical pressure coefficient of
an electron-doped high-T$_{c}$ superconductor with a similar ambient
pressure T$_{c}$.

In the discussion of the pressure effects the lattice anisotropy has to be
considered since even hydrostatic pressure may result in different
compression ratios along the main crystallographic directions. In optimally
doped YBCO, for example, the dT$_{c}$/dp varies in sign and magnitude for
compression along a-, b-, and c-axis resulting in a very small overall
pressure shift of T$_{c}$ under hydrostatic pressure conditions.\cite{7} The
lattice anisotropy in Na$_{0.35}$CoO$_{2}\cdot $yH$_{2}$O is significant due
to the huge expansion of the c-axis after intercalation of water and the
compression is expected to be anisotropic as well. High pressure x-ray
investigations can distinguish the compression of a- and c-axis and are
therefore highly desirable. For the current discussion we assume that the
two-dimensionality and the expansion of the c-axis by intercalation of water
in Na$_{0.35}$CoO$_{2}\cdot $yH$_{2}$O is the key to understand the
superconductivity (this is at least assumed in all previous publications).%
\cite{1,2,3,4} Then it is naturally to expect that a pressure induced
decrease of the c-axis length results in a suppression of the
superconducting state. However, the effects of in-plane compression are not
clear and need to be investigated separately. Uniaxial pressure experiments
using single crystals of Na$_{0.35}$CoO$_{2}\cdot $yH$_{2}$O (not currently
available) can help to separate the different contributions. A pressure
induced charge transfer to the CoO$_{2}$-layers (as discussed for the
cuprate superconductors) is unlikely as long as the intercalated water
molecules remain neutral and the valence of the Na-ions is +1. The
non-linear decrease of T$_{c}$ observed in our investigation is equally
unusual and interesting. It may suggest the possible existence of large
dispersion in the energy spectrum of the compound near its Fermi surface.

In conclusion, we have investigated the dependence of the superconducting T$%
_{c}$ of Na$_{0.35}$CoO$_{2}\cdot $yH$_{2}$O upon hydrostatic pressure up to
1.6 GPa. The pressure coefficient is negative and T$_{c}$(p) shows a strong
nonlinearity. The sign and magnitude of the relative pressure coefficient
are compatible with similar data of electron-doped cuprates. Our results
indicate that the superconductivity in this compound may originate from the
apparent two-dimensionality of the CoO$_{2}$ planes similar to the CuO$_{2}$
layers of the electron-doped copper oxides.

\begin{acknowledgments}
This work is supported in part by NSF Grant No. DMR-9804325, the T.L.L.
Temple Foundation, the John J. and Rebecca Moores Endowment, and the State
of Texas through the TCSAM at the University of Houston and at Lawrence
Berkeley Laboratory by the Director, Office of Energy Research, Office of
Basic Energy Sciences, Division of Materials Sciences of the U.S. Department
of Energy under Contract No. DE-AC03-76SF00098.
\end{acknowledgments}

\begin{figure}[tbp]
\caption{ac susceptibility of Na$_{0.35}$CoO$_{2}\cdot $yH$_{2}$O measured
at different pressures. The data are normalized to the susceptibility value
at 5 K. The superconducting transition temperature was estimated from the
onset of the diamagnetic signal, as indicated by the arrow in the inset. The
sequence of pressure application follows the values listed in the figure.
The filled squares denote the data set acquired after pressure was released.}
\label{F1}
\end{figure}

\begin{figure}[tbp]
\caption{Pressure dependence of $T_{c}$ of Na$_{0.35}$CoO$_{2}\cdot $yH$_{2}$%
O. The open circles are data taken at increasing pressure, the filled square
indicates T$_{c}$ after the pressure was completely released. The numbers
indicate the chronological order of the pressure changes.}
\label{F2}
\end{figure}


\begin{references}
\bibitem{1}  K. Takada, H. Sakurai, E. Takayama-Muromachi, F. Izumi, R. A. 
Dilanian, T. Sasaki, Nature 422, 53 (2003).
\bibitem{2} G. Baskaran, cond-mat/0303649, unpublished.
\bibitem{3} Q.-H. Wang, D.-H. Lee, P. A. Lee, cond-mat/0304377, unpublished.
\bibitem{4} A. Tanaka, X. Hu, cond-mat/0304409, unpublished.
\bibitem{5} J. Cmaidalka et al., to be published.
\bibitem{9} T. F. Smith, C. W. Chu, Phys. Rev. 159, 353 (1967).
\bibitem{6} C. W. Chu, L. R. Testardi, Phys. Rev. Lett. 32, 766 (1974).
\bibitem{10} B. Lorenz, R. L. Meng, C. W. Chu, Phys. Rev. B 64, 012507 (2001).
\bibitem{7} J. S. Schilling, S. Klotz, Physical Properties
of High Temperature Superconductors III, Edt. D. Ginsberg, World Scientific
New York (1992), p. 59.
\bibitem{8} J. T. Markert, J. Beille, J. J. Neumeier, E. A. Early, C. L. Seaman,
T. Moran, M. B. Maple, Phys. Rev. Letters 64, 80 (1990).
\end{references}

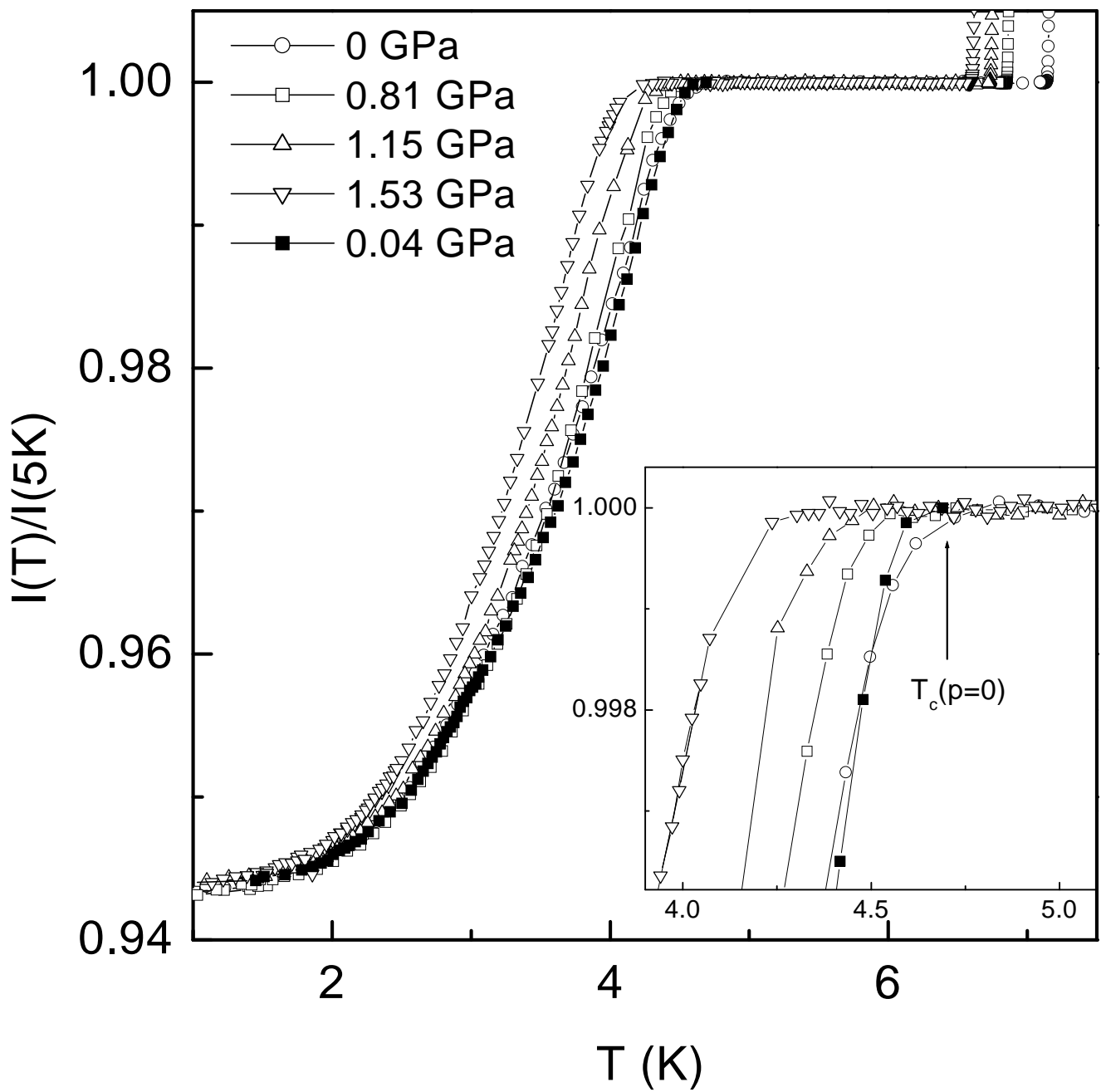
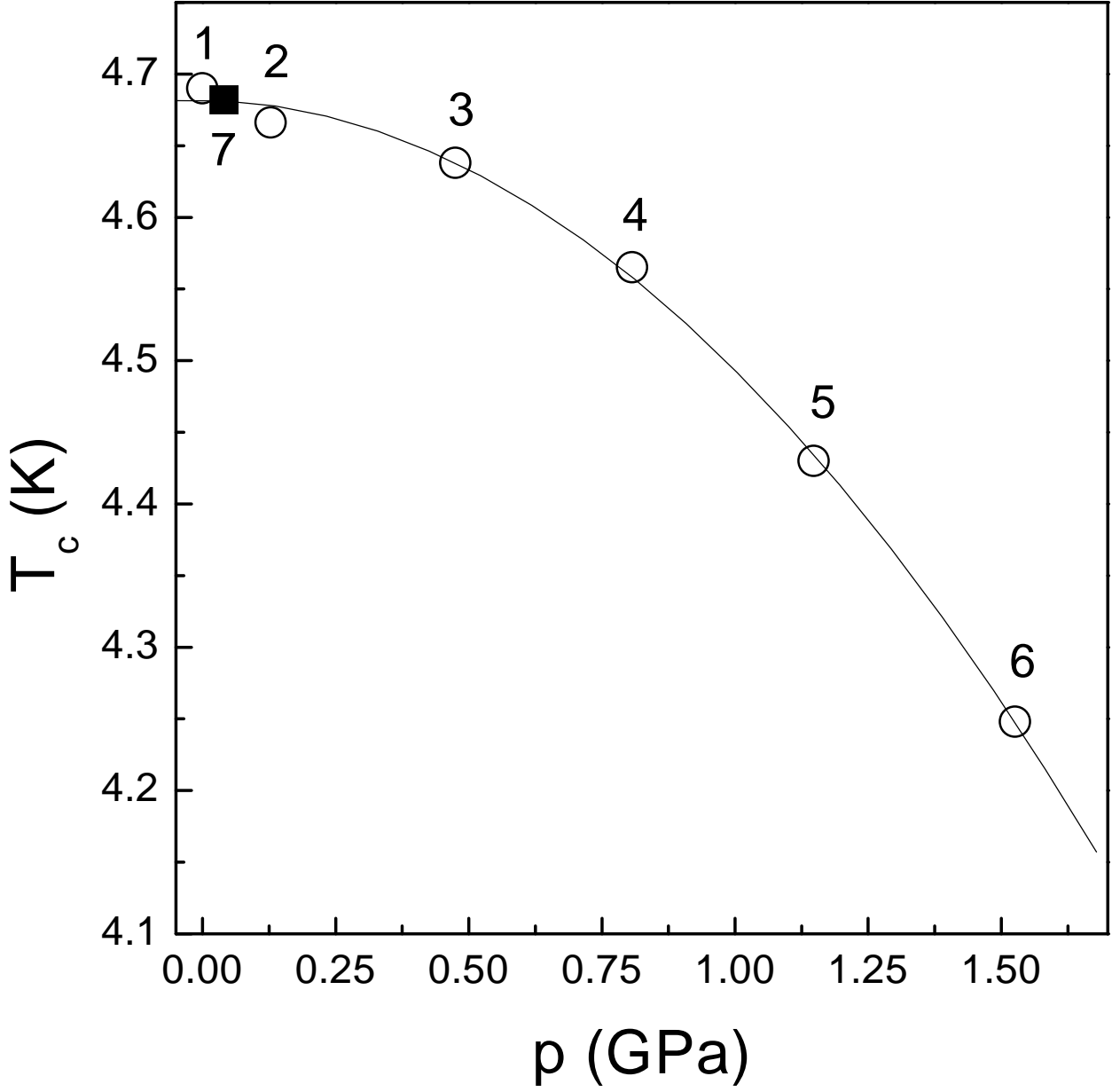
\end{document}